\documentclass[sigconf]{acmart}

\usepackage{tabularx}
\usepackage{multirow}
\usepackage{array}
\usepackage{bm}
\usepackage{hyperref}
\usepackage{cleveref}
\crefname{section}{Section}{Sections}
\usepackage{enumitem}
\setlist[itemize]{left=0pt, topsep=0pt}

\usepackage[linesnumbered,ruled,vlined]{algorithm2e}
\SetAlFnt{\small}
\SetAlCapFnt{\small}

\renewcommand\footnotetextcopyrightpermission[1]{} 

\usepackage{listings, xcolor}
\definecolor{lightgreen}{rgb}{.12,.82,.31}

\lstdefinelanguage{Solidity}{
	keywords=[1]{anonymous, assembly, assert, balance, break, call, callcode, case, catch, class, constant, continue, constructor, contract, debugger, default, delegatecall, delete, do, else, emit, event, experimental, export, external, false, finally, for, function, gas, if, implements, import, in, indexed, instanceof, interface, internal, is, length, library, log0, log1, log2, log3, log4, memory, modifier, new, payable, pragma, private, protected, public, pure, push, require, return, returns, revert, selfdestruct, send, Solidity, storage, struct, suicide, super, switch, then, this, throw, transfer, true, try, typeof, using, value, view, while, with, addmod, ecrecover, keccak256, mulmod, ripemd160, sha256, sha3}, 
	keywordstyle=[1]\color{blue}\bfseries,
	keywords=[2]{address, bool, byte, bytes, bytes1, bytes2, bytes3, bytes4, bytes5, bytes6, bytes7, bytes8, bytes9, bytes10, bytes11, bytes12, bytes13, bytes14, bytes15, bytes16, bytes17, bytes18, bytes19, bytes20, bytes21, bytes22, bytes23, bytes24, bytes25, bytes26, bytes27, bytes28, bytes29, bytes30, bytes31, bytes32, enum, int, int8, int16, int24, int32, int40, int48, int56, int64, int72, int80, int88, int96, int104, int112, int120, int128, int136, int144, int152, int160, int168, int176, int184, int192, int200, int208, int216, int224, int232, int240, int248, int256, mapping, string, uint, uint8, uint16, uint24, uint32, uint40, uint48, uint56, uint64, uint72, uint80, uint88, uint96, uint104, uint112, uint120, uint128, uint136, uint144, uint152, uint160, uint168, uint176, uint184, uint192, uint200, uint208, uint216, uint224, uint232, uint240, uint248, uint256, var, void, ether, finney, szabo, wei, days, hours, minutes, seconds, weeks, years},	
	keywordstyle=[2]\color{teal}\bfseries,
	keywords=[3]{block, blockhash, coinbase, difficulty, gaslimit, number, timestamp, msg, data, gas, sender, sig, value, now, tx, gasprice, origin},	
	keywordstyle=[3]\color{violet}\bfseries,
	identifierstyle=\color{black},
	sensitive=true,
	comment=[l]{//},
	morecomment=[s]{/*}{*/},
	commentstyle=\color{gray}\ttfamily,
	stringstyle=\color{lightgreen}\ttfamily,
	morestring=[b]',
	morestring=[b]"
}

\begin{document}

\title[CKG-LLM: Detection of Smart Contract Access Control Vulnerabilities]{CKG-LLM: LLM-Assisted Detection of Smart Contract Access Control Vulnerabilities Based on Knowledge Graphs}

\author{Xiaoqi Li}
\affiliation{%
  \institution{Hainan University}
  \city{Haikou}
  \country{China}}
\email{csxqli@ieee.org}

\author{Hailu Kuang}
\authornote{The corresponding author.}
\affiliation{%
  \institution{Hainan University}
  \city{Haikou}
  \country{China}}
\email{hailukuang@hainanu.edu.cn}

\author{Wenkai Li}
\affiliation{%
  \institution{Hainan University}
  \city{Haikou}
  \country{China}}
\email{cswkli@hainanu.edu.cn}

\author{Zongwei Li}
\affiliation{%
  \institution{Hainan University}
  \city{Haikou}
  \country{China}}
\email{lizw1017@hainanu.edu.cn}

\author{Shipeng Ye}
\affiliation{%
  \institution{Hainan University}
  \city{Haikou}
  \country{China}}
\email{shipengye@hainanu.edu.cn}

\renewcommand{\shortauthors}{Xiaoqi Li, Hailu Kuang, et al.}

\begin{abstract}
Traditional approaches for smart contract analysis often rely on intermediate representations such as abstract syntax trees, control-flow graphs, or static single assignment form. However, these methods face limitations in capturing both semantic structures and control logic. Knowledge graphs, by contrast, offer a structured representation of entities and relations, enabling richer intermediate abstractions of contract code and supporting the use of graph query languages to identify rule-violating elements.
This paper presents CKG-LLM, a framework for detecting access-control vulnerabilities in smart contracts. Leveraging the reasoning and code generation capabilities of large language models, CKG-LLM translates natural-language vulnerability patterns into executable queries over contract knowledge graphs to automatically locate vulnerable code elements. Experimental evaluation demonstrates that CKG-LLM achieves superior performance in detecting access-control vulnerabilities compared to existing tools. Finally, we discuss potential extensions of CKG-LLM as part of future research directions.

\end{abstract}

\ccsdesc[500]{Security and privacy~Software security engineering}

\keywords{Smart Contract, LLM, KG, Blockchain Security, Access Control}

\maketitle

\section{Introduction}

Smart contracts are programs stored on a blockchain that automatically execute predefined rules \cite{szabo1996smart, gao2025implementation, shen2025blockchain, liu2025empirical, huang2025comparative}. Knowledge graphs (KGs) \cite{hogan2021knowledge} can not only be used to represent entities and connections in the real world, but also have the potential to serve as an intermediate representation of the semantic structure and execution logic of smart contract code.
Research conducted by Hu et al. \cite{hu2023detect} demonstrates the feasibility of constructing a smart contract knowledge graph (CKG) by parsing abstract syntax trees (ASTs) \cite{neamtiu2005understanding} of smart contracts and utilizing graph query languages (GQL) to identify contract vulnerabilities.

However, the contract knowledge graph extracted solely from AST may result in shallow semantic representations and fail to capture deep syntactic and structural information \cite{peng2025multicfv}. Moreover, the difficulty in conceiving and writing SPARQL statements that accurately locate entities or relationships will limit the promotion of knowledge graph-based detection methods in contract security.

To address these challenges, we propose CKG-LLM, a LLM-assisted framework for detecting access control vulnerabilities based on smart contract knowledge graphs.
The framework exploits Slither's intermediate representation (IR) \cite{feist2019slither} to extract diverse instances, containment relations, execution semantics, control-flow structures, and invocation dependencies, thereby constructing a semantically enriched CKG.
Furthermore, the framework leverages the code generation and reasoning capabilities of large language models (LLMs) \cite{chang2024survey, xiang2025security} to translate natural-language (NL) vulnerability patterns into graph queries, thereby eliminating the need for rule-based parsing.
To enhance query accuracy and robustness, we further introduce RLAF, a domain-adaptive reinforcement learning approach that incorporates feedback from an auxiliary language model agent, enabling iterative refinement of query generation.

The main contributions of this paper are summarized as follows:
\begin{itemize}
    \item We propose CKG-LLM, to the best of our knowledge, the first framework that leverages LLMs to translate natural language descriptions into graph queries over smart contract knowledge graphs, enabling access control vulnerability detection.
    \item We develop a contract knowledge graph construction pipeline based on Slither's intermediate representation, and introduce a reinforcement learning from small language model agent feedback (RLAF) to guide LLMs in generating high-quality queries.
    \item Through evaluation, CKG-LLM achieved an average accuracy of 73.1\% and an F1 score of 74.9\%,  demonstrating higher detection efficiency compared to existing access-control analysis tools.
\end{itemize}

In future work, we plan to extend CKG-LLM to cover additional classes of vulnerabilities and refine its pipeline by integrating retrieval-augmented generation (RAG) \cite{lewis2020retrieval} for SPARQL query generation, thereby improving both scalability and accuracy.

\section{Background}

In recent years, Access control has climbed from fifth to first place in the OWASP rankings, indicating its increasing prevalence and severity in real-world software development \cite{owasp2025, wu2025security, li2025atomgraph, luo2025movescanner}. Access control vulnerabilities are particularly critical among various vulnerabilities in smart contracts. As contracts are immutable and autonomously manage digital assets, any flaw in authorization logic can allow attackers to gain unauthorized control, resulting in significant and irreversible asset loss \cite{li2025beyond, peng2025mining, zhang2025security, li2025penetrating}. Therefore, accurately detecting access control vulnerabilities is essential for ensuring contract security. 

Recent work demonstrates that LLMs can strengthen smart contract security pipelines. LLMs can complement traditional techniques such as static analysis, symbolic execution, and fuzz testing \cite{sun2024gptscan, chen2025numscout, shou2024llm4fuzz, zhang2025risk, wang2025ai}. 
Instead of relying on traditional program analysis techniques, the KG-based contract detection organizes code elements and their relations into graph-based representations. Researchers have demonstrated that by mapping syntactic and structural aspects into ontological and instance-level entities, contract logic can be captured in a semantically enriched graph form \cite{hu2023detect}.

LLMs have gained widespread adoption in translating natural language into query languages (NL2GQL) \cite{zhou2023r} \cite{liang2024aligning}, which lowers the barrier for interacting with knowledge graphs. Within the domain of smart contracts, users can describe security needs or potential vulnerability patterns in natural language, and LLMs automatically generate corresponding GQL queries to detect contract flaws. Combining LLMs with KG-based contract representations enhances domain-specific knowledge, addresses reasoning limitations, and enables the detection of identity-related vulnerabilities.

\section{CKG-LLM}
\label{sec::overview}
\begin{figure*}[htbp]
    \centering
    \includegraphics[width=0.9\textwidth, trim=2 30 2 20, clip]{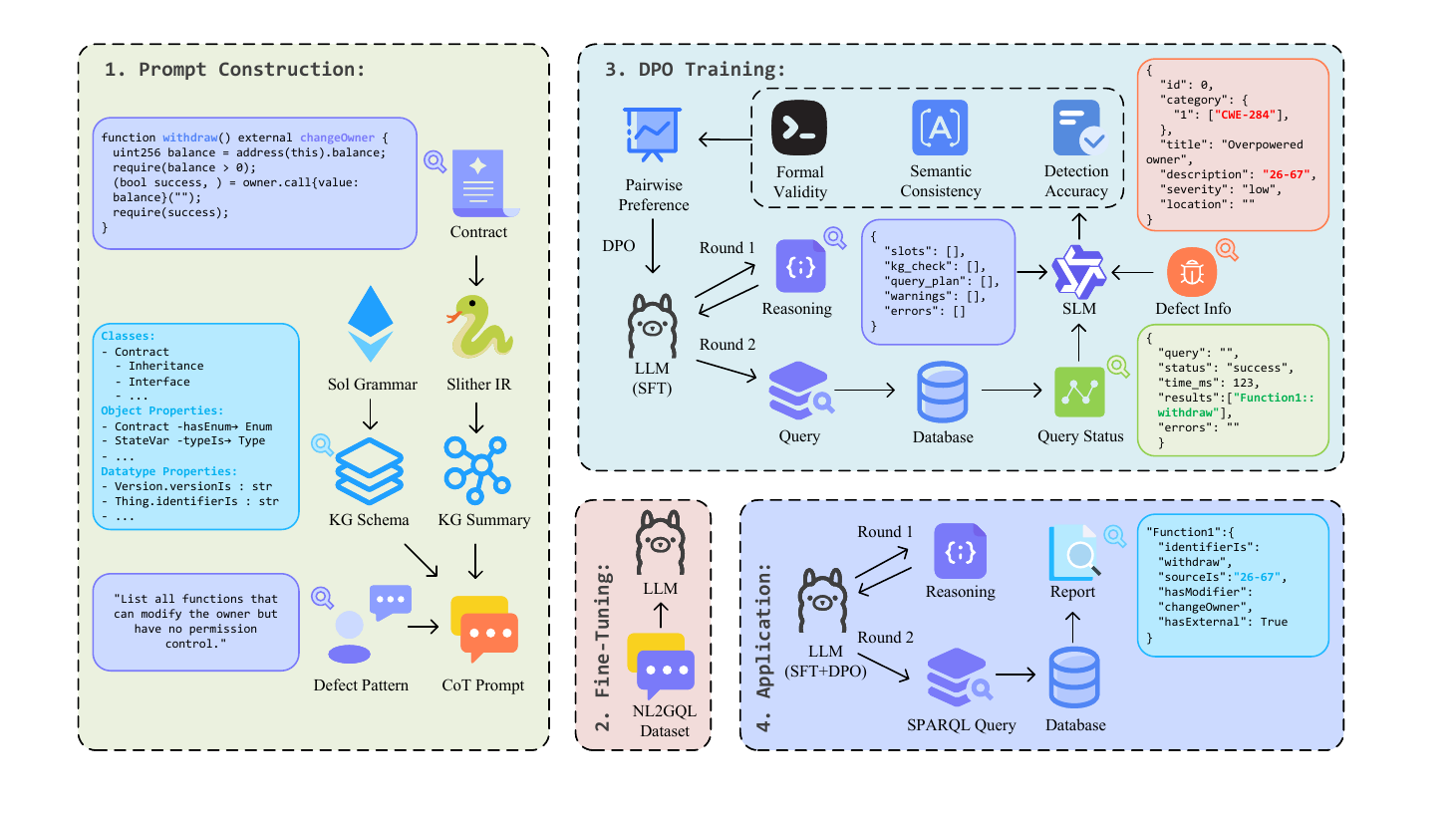}
    \caption{The overall workflow of CKG-LLM, consists of four stages: (1) construction of the contract knowledge graph and the prompt design; (2) supervised fine-tuning for guiding LLMs in NL2GQL tasks; (3) DPO-based training to enhance the effectiveness of generated GQL in detecting vulnerabilities; and (4) LLM-assisted query generation for real-world applications.}
    \Description{The overall workflow of CKG-LLM.}
    \label{fig:CKG-LLM_overview}
\end{figure*}

CKG-LLM is a graph-query-driven framework for smart contract vulnerability detection. It employs domain-enhanced fine-tuning and reinforcement learning (RL) to guide LLMs in generating GQL queries, thereby enabling the identification of vulnerabilities within contract knowledge graphs.
The overall workflow, shown in Figure~\ref{fig:CKG-LLM_overview}, comprises four stages: (1) contract knowledge graph construction and prompt design; (2) supervised fine-tuning for NL2GQL; (3) DPO-based reinforcement learning from small language model agent feedback (RLAF); and (4) framework deployment and application. Each stage is described in the following sections.

\subsection{CKG Building}
The CKG employs a two-layered design, comprising the ontology and the instance layer. The ontology layer formalizes the abstract syntax of Solidity, serving as a domain schema, while the instance layer represents contract structures and their interrelations. 
In this paper, KG classes are written in Small Caps (e.g., \textsc{Contract}), properties are presented in Monospace (e.g., \texttt{hasFunction}), and instances are italicized (e.g., \textit{Contract1}).

\subsubsection{Ontology}
\begin{figure}[htbp]
    \centering
    \includegraphics[width=0.5\textwidth, trim=24 20 2 28, clip]{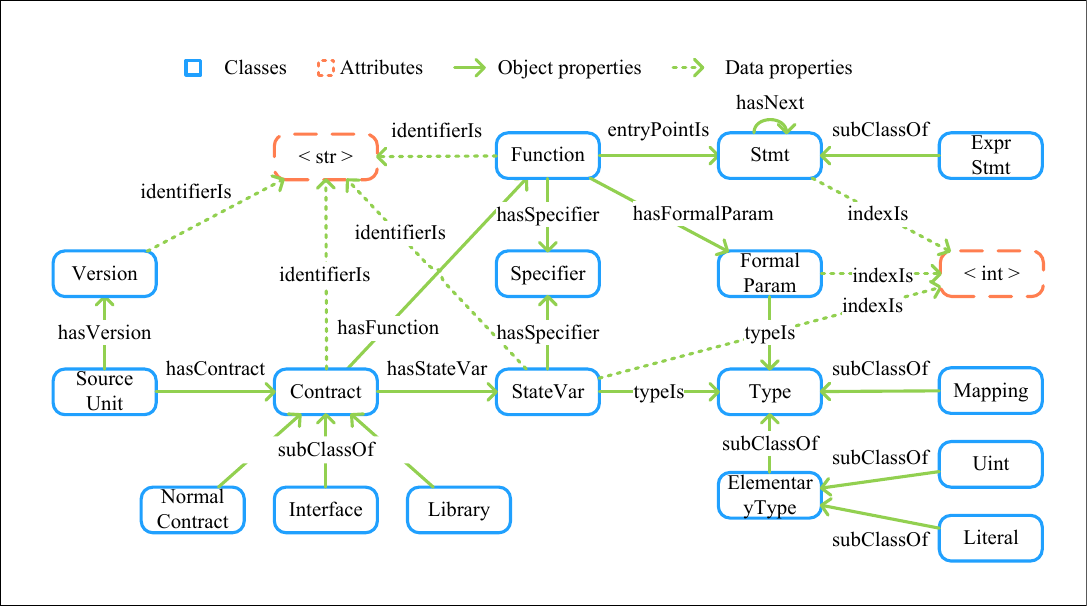}
    \caption{A simplified example of the CKG ontology layer, where nodes represent classes and their attributes, while edges denote object and instance properties.}
    \Description{A simplified example of the CKG ontology layer.}
    \label{fig:ontology_layer}
\end{figure}

The ontology layer construction is informed by the Solidity language grammar \cite{soliditygrammar2025} and the ANTLR grammar specification \cite{antlrsolidity2025}. Formally, we define the ontology as:
\begin{displaymath}
\mathcal{O} = (C, A, P),
\end{displaymath}
where $C$ denotes the set of ontology classes,  $A$ the set of class attributes, and  $P = P_{\text{obj}} \cup P_{\text{data}}$ the set of properties, with $P_{\text{obj}} \subseteq C \times P \times C$ representing object properties that connect classes, and $P_{\text{data}} \subseteq C \times P \times A$ representing datatype properties that link classes to their attributes. As shown in Figure~\ref{fig:ontology_layer}, ontology class nodes encompass syntactic constructs, including \textsc{Contracts}, \textsc{Functions}, \textsc{Statements}, and \textsc{StateVar}. 
These classes may also form inheritance-like relationships (e.g., \textsc{Interface} as a subclass of \textsc{Contract}). Class attribute nodes define the data types (e.g., the \textsc{Identifier} data type is string). Properties between nodes are further categorised as \texttt{has} or \texttt{is}, depending on their cardinality (e.g., \textsc{Contract} \texttt{hasStateVar} \textsc{StateVar}).
We also introduce auxiliary nodes and edges to capture program logic and control flow (e.g., \texttt{hasNext} encodes execution order), facilitate efficient querying (e.g., \texttt{indexIs} specifies the position of statements or struct members), or improve interpretability to LLMs (e.g., \texttt{entryPointIs} links a function to its entry statement).

\subsubsection{Instance}
\begin{figure}[htbp]
    \centering
    \includegraphics[width=0.4\textwidth, trim=24 20 2 24, clip]{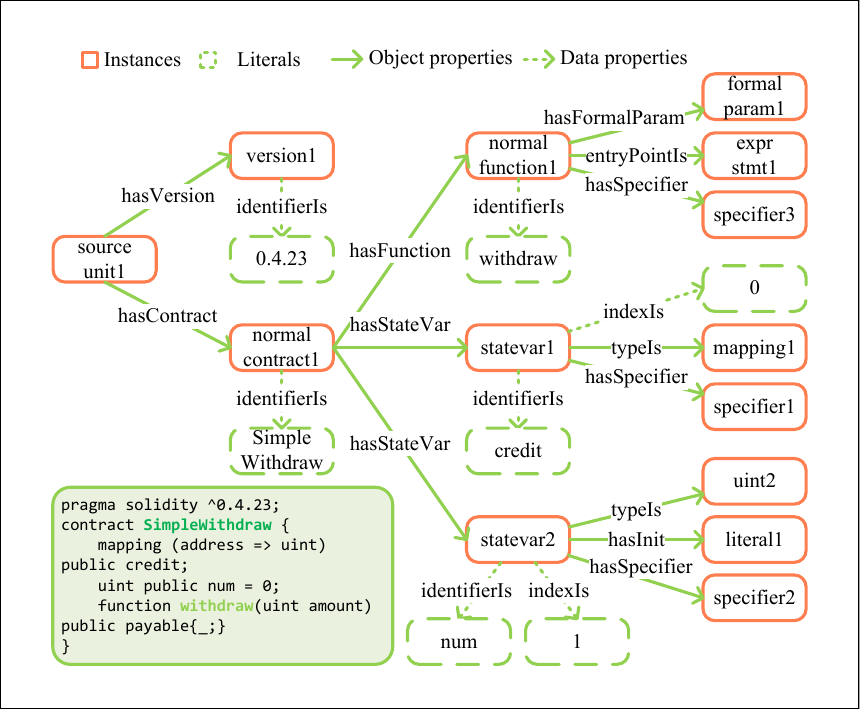}
    \caption{A simplified example of the CKG instance layer for the SimpleWithdraw contract, detailing instances and literal value nodes, along with object and data property edges.}
    \label{fig:instance_layer}
\end{figure}

The instance layer is constructed using Slither \cite{feist2019slither}, which functions not only as a static analysis tool but also as a contract parser. 
Formally, the instance layer is represented as:
\begin{displaymath}
\mathcal{I} = (I, L, P),
\end{displaymath}
where $I$ denotes the set of instance nodes,  
$L$ the set of literal values associated with specific instances, and  
$P = P_{\text{obj}}^I \cup P_{\text{data}}^I$ the set of instantiated properties,  
with $P_{\text{obj}}^I \subseteq I \times P \times I$ denoting object property assertions 
that connect instances, and $P_{\text{data}}^I \subseteq I \times P \times L$ 
denoting datatype property assertions that connect instances to their literal values.
In our method, Slither compiles the contract to extract its AST and decomposes statements into operations to build Slither IR. This IR is further converted into an SSA form by assigning each variable a unique version and adding $\phi$-functions to handle control-flow divergence in variable values, making data and control flows explicit for richer analysis. 
As shown in Figure~\ref{fig:instance_layer}, by traversing the Slither IR, we can extract not only contract-level elements as nodes and relations as properties for the instance layer, but also more complex relationships, including contract or function \textit{Inheritance}, function or modifier \textit{Invocation}, and statement-level execution flows that are often difficult to obtain directly from parsing the ASTs.

\subsection{Prompt Construction}
\label{sec::prompt}
\begin{figure}[htbp]
    \centering
    \includegraphics[width=0.45\textwidth, trim=18 16 2 22, clip]{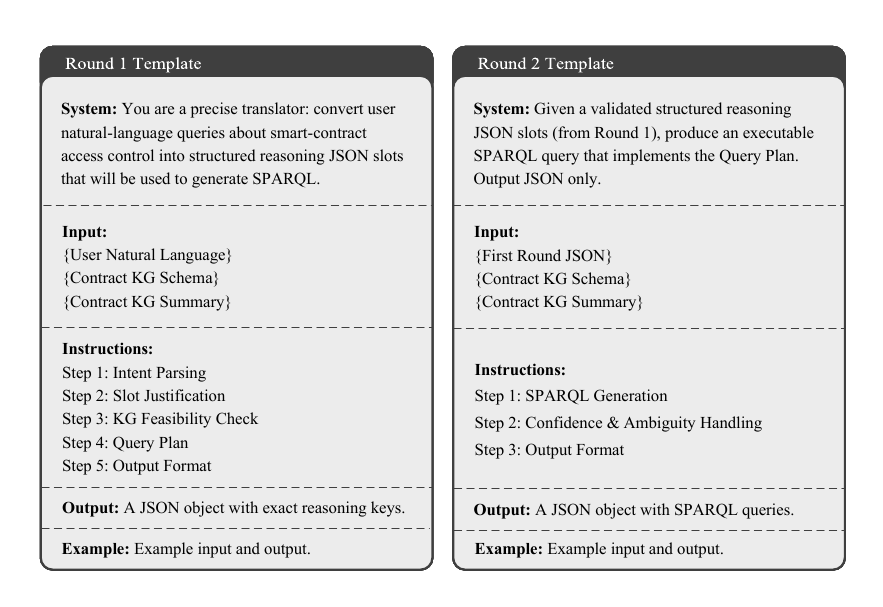}
    \caption{An overview of the two-round prompt template. The first round requires intermediate reasoning, and the second round requires queries based on the reasoning results.}
    \Description{An overview of the two-round prompt template.}
    \label{fig:prompt_template}
\end{figure}

We design a two-round chain of thought (CoT) prompts \cite{wei2022chain} to enable LLMs to translate natural-language defect patterns into SPARQL queries over the contract knowledge graph. The illustration of the prompt template is shown in Figure~\ref{fig:prompt_template}. Each round follows a unified template with five components: Task, Input, Instructions, Output, and Example. 
The first round involves guiding structured reasoning from natural language, whereas the second round generates executable SPARQL queries based on the reasoning output. Compared to conventional NL2GQL prompts, our design modifies the Input and Instructions sections better to align query generation with the contract knowledge graph.

\subsubsection{Input}
The prompt input incorporates a structured CKG schema, enabling the model to interpret domain-specific terminology and relation mappings, as well as a summary of the KG in tabular form, to facilitate feasibility checks and constrain query generation within the boundaries of the CKG.
The KG summary is extracted through a pruning strategy that extracts only access-control-related subgraphs to avoid limitations in the LLM context. 
As described in Algorithm \ref{alg:ac-pruning}, 
It first identifies public or external entry functions, then verifies the presence of access-control checks (e.g., \textit{Modifiers}), and finally retains variables relevant to authority (e.g., \textit{owner}). The resulting KG summary is reconstructed from the pruned triples and serves as the compact input for query generation.

\begin{algorithm}[htbp]
\caption{Pruning Contract KG for Access Control}
\label{alg:ac-pruning}
\DontPrintSemicolon
\SetKwInOut{Input}{Input}\SetKwInOut{Output}{Output}

\Input{Contract knowledge graph $G=(V,E)$}
\Output{Pruned subgraph $G' \subseteq G$}

$R \gets \emptyset$  \tcp*[r]{Relevant nodes}

\ForEach{function $f \in V$}{
    \If{$visibility(f) \in \{\text{public}, \text{external}\}$}{
        \If{$f$ has modifier $\in \{\text{onlyOwner}, \text{onlyRole}\}$
            or contains statement involving 
            $\{\text{msg.sender}, \text{require}, \text{hasRole}\}$}{
                $R \gets R \cup \{f\}$\;
                $R \gets R \cup$ variables 
                    $\{v \mid v \in \{\text{owner}, \text{roles}\}\}$ used in $f$\;
        }
    }
}

$G' \gets \{(s,p,o) \in E \mid s \in R \lor o \in R\}$\;

\Return{$G'$}

\end{algorithm}

\subsubsection{Instructions}
The prompt instruction serves as the core, guiding the LLM through a structured CoT process for NL2GQL generation. We decompose the NL2GQL task into seven essential instructions: Intent Parsing, Slot Justification, KG Feasibility Check, Query Plan Construction, SPARQL Generation, Confidence and Ambiguity Handling, and Output Formatting.
Intent Parsing uses the LLM’s semantic understanding to normalize target entities and conditions according to the Contract\_KG\_Schema, while Slot Justification links natural-language triggers to their corresponding schema fields. KG Feasibility Check validates whether slot values can be mapped to ontology or instance elements and reports mismatches. Query Plan Construction extracts minimal triple patterns and constraints, which SPARQL Generation converts into ontology-consistent queries. The Confidence and Ambiguity Handling class categorizes outputs by confidence and provides alternatives for ambiguous cases, while the Output Formatting standardizes the final query structure for consistency.

\subsection{Fine-Tuning}
\label{sec::sft}
Supervised fine-tuning (SFT) is designed to familiarize the model with GQL paradigms. In our framework, we choose Llama-3.1-8B-Instruct \cite{llama} as the base model due to its code generation capability and open-source nature. Regarding dataset choice, we utilize LC-QuAD 2.0 \cite{lcquad2}, which comprises a collection of 55,782 complex query pairs, aligning well with the intricate query requirements of smart contract knowledge graphs. We also adopt the low-rank adaptation (LoRA) method \cite{hu2022lora} to reduce computational cost and memory usage while preserving model expressiveness.

\subsection{DPO Training}
\label{sec::DPO}
To enhance the LLM's understanding of contract syntax and vulnerability detection rules, we propose a reinforcement learning approach that leverages feedback from a small language model agent (RLAF). Specifically, we input the 1,291 contracts from the FORGE dataset \cite{chen2025forge} (refer to \cref{sec::exp}) along with simplified Access Control-related CWE patterns \cite{CWE-284, CWE-285, CWE-863, CWE-862, CWE-269, CWE-276, CWE-732} into the module described in Stage~\hyperref[sec::prompt]{1} to construct the LLM prompt. These prompts then served as inputs for the model trained in Stage~\hyperref[sec::sft]{2}. Each prompt undergoes two rounds of dialogue to generate a single SPARQL query, and each prompt is executed twice to construct query pairs for comparison. The small language model (SLM) is a specialized Qwen2.5-3B-Instruct \cite{qwen} that evaluates LLM-generated reasoning results, SPARQL queries, query status from the graph database, and defect information from the dataset to rank query pairs for direct preference optimization (DPO) training \cite{rafailov2023direct, li2025facial}.
The evaluation is based on three dimensions: Formal Validity (whether the query executes correctly and retrieves results relevant to the NL intent), Semantic Consistency (whether the query semantically aligns with the intended meaning, respects the schema, and uses compatible filtering conditions such as visibility or call relations), and Detection Accuracy (how effectively the query detects vulnerabilities, measured via localization accuracy, type coverage, and the precision of retrieved instances).

\subsection{Application}
\label{sec::app}
In real-world contract detection, the framework takes as input both user-defined natural language vulnerability descriptions and the source code of smart contracts, feeding them into the models discussed in Stage~\hyperref[sec::prompt]{1}. It proceeds to construct a two-round domain-enhanced chain of thought prompt, which is then fed into the pretrained LLM to generate a SPARQL query tailored for vulnerability localization. The generated query is executed against a contract knowledge graph stored in a graph database, yielding the target entities, and proceeds to create final detection results.

\section{Experimental Evaluation}
\label{sec::exp}

In this section, we will address the following research questions:
\textbf{\underline{RQ1 (Quality):}} How is the quality of LLM generated queries?
\textbf{\underline{RQ2 (Accuracy):}} Can SPARQL query-based detection accurately identify access control vulnerabilities in real-world smart contracts?
\textbf{\underline{RQ3 (Efficiency):}} How efficient is CKG-LLM compared with existing frameworks for access control vulnerability detection?

Our experiments were conducted on a server equipped with an Intel i7-13700KF CPU, 32GB of RAM, and an NVIDIA RTX 4070 GPU, running on Ubuntu 24.04. The experimental dataset was sourced from FORGE \cite{chen2025forge}, from which we filtered 1,844 contracts that contained defect labels according to seven access control-related CWE patterns \cite{CWE-284, CWE-285, CWE-863, CWE-862, CWE-269, CWE-276, CWE-732}. Among them, 1,291 contracts were utilized for DPO training in \cref{sec::DPO}, and 553 contracts for evaluation.

\begin{figure}[htbp]
    \centering
    \includegraphics[width=0.4\textwidth, trim=10 28 2 10, clip]{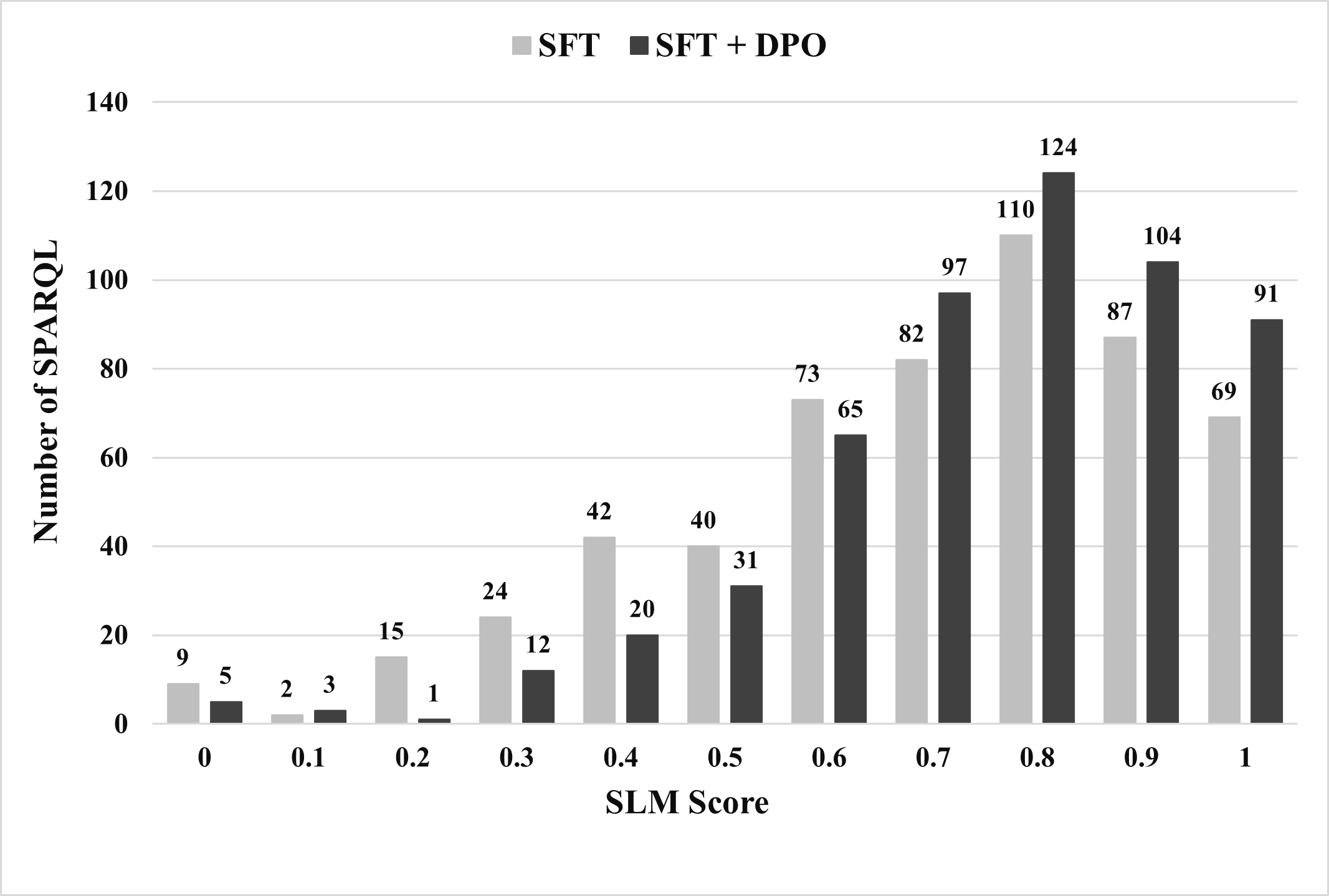}
    \vspace{-1ex}
    \caption{The SLM score distribution of the SFT and SFT+DPO trained model over 553 contracts in NL2GQL tasks.}
    \vspace{-3ex}
    \Description{The SLM score distribution.}
    \label{fig:distribution}
\end{figure}

\textbf{RQ1 Quality:}
To address RQ1, we reused the training pipeline from Stage \hyperref[sec::DPO]{3} to construct the evaluation workflow. Both the SFT and the SFT+DPO trained LLMs were employed to assess their effectiveness on the NL2GQL task. Similar to the DPO training setup, the evaluation inputs consisted of 553 contracts with access-control-related defect labels from the FORGE subset \cite{chen2025forge}, along with the defect pattern derived from the concise CWE-284 description \cite{CWE-284}.
In this experiment, we made a minor modification to the prompting strategy. For each contract, the LLM was asked to execute the prompt once to produce a single query, while the SLM was asked to provide scoring rather than ranking of generated results. The average generation quality was then computed across all 553 contracts, with the score distribution presented in Figure~\ref{fig:distribution}.
The results in Figure~\ref{fig:distribution} demonstrate that both models achieved relatively high scores in the NL2GQL task, while incorporating DPO into SFT improves the quality of SPARQL generation. The SFT model achieved an average score of 0.693, and the reinforcement-learned models scored even higher, at 0.756. This shift indicates that DPO enhances the model's ability to generate more accurate and reliable queries.

\textbf{RQ2 Accuracy:}
To address RQ2, we assessed the detection accuracy and F1 scores of the Llama base, SFT, and SFT+DPO models across 553 contracts against the CWE-284 pattern. We followed the setup in Stage~\hyperref[sec::app]{4} and compared their query outputs against the dataset labels with the assistance of SLM. The statistical results are summarized in Table~\ref{tab:f1_score}.
As shown in the table, SFT training yields a notable improvement in precision. It reduces recall compared to the baseline, demonstrating the effectiveness of supervised fine-tuning in aligning the model with the NL2GQL task. Furthermore, the DPO training yields consistent gains across all metrics, indicating that preference optimization not only reduces false positives but also enhances the overall robustness of vulnerability detection.

\begin{table}[ht]
    \small
    \centering
    \caption{Detection accuracy comparison of baseline, SFT, and SFT+DPO models. Metrics: true positives (TP), false positives (FP), false negatives (FN), precision, recall, and F1 score.}
    \Description{Detection accuracy comparison results.}
    \label{tab:f1_score}
    \begin{tabular}{>{\centering\arraybackslash}m{1cm} >{\centering\arraybackslash}m{0.6cm} >{\centering\arraybackslash}m{0.6cm} >{\centering\arraybackslash}m{0.6cm} >{\centering\arraybackslash}m{0.8cm} >{\centering\arraybackslash}m{0.8cm} >{\centering\arraybackslash}m{0.8cm}} 
        \toprule
        \textbf{Model} & \textbf{\#TP} & \textbf{\#FP} & \textbf{\#FN} & \textbf{Prec(\%)} & \textbf{Recall(\%)} & \textbf{F1(\%)}\\
        \midrule
        Base & 276 & 223 & 180 & 55.3 & 60.5 & 57.8\\
        \midrule
        SFT & 314 & 197 & 134 & 61.4 & 70.1 & 65.5\\
        \midrule
        \textbf{DPO+SFT} & \textbf{348} & \textbf{128} & \textbf{105} & \textbf{73.1} & \textbf{76.8} & \textbf{74.9}\\
        \bottomrule
    \end{tabular}
\end{table}

\textbf{RQ3 Efficiency:}
To address RQ3, we measured the total detection time of CKG-LLM, covering the entire pipeline of prompt construction, LLM response, and report generation based on the accuracy validation process in \cref{sec::exp}. For comparison, we further introduced Achecker \cite{ghaleb2023achecker}, an access control vulnerability detection framework, and GPTScan \cite{sun2024gptscan}, an LLM-assisted static analysis framework, to analyze both accuracy and efficiency. The results are presented in Table \ref{tab:comparison}.
The comparative results highlight that CKG-LLM offers the most balanced performance in terms of both accuracy and efficiency. Although Achecker achieves the highest F1 score (81.4\%), this advantage comes at the cost of higher runtime. GPTScan, on the other hand, performs competitively but falls short in both accuracy and efficiency. In contrast, CKG-LLM achieves a strong F1 score (74.9\%) while delivering the lowest overall runtime and the fastest average detection time per contract.

\begin{table}[ht]
    \small
    \centering
    \caption{Detection accuracy and efficiency comparison of Achecker, GPTScan, and CKG-LLM. Metrics: precision, recall, F1 score, total detection time, and average detection time.}
    \Description{Detection efficiency comparison results.}
    \label{tab:comparison}
    \begin{tabular}{>{\centering\arraybackslash}m{1.4cm} >{\centering\arraybackslash}m{0.8cm} >{\centering\arraybackslash}m{0.8cm} >{\centering\arraybackslash}m{0.8cm} >{\centering\arraybackslash}m{0.8cm} >{\centering\arraybackslash}m{0.8cm}} 
        \toprule
        \textbf{Tool} & \textbf{Prec(\%)} & \textbf{Recall(\%)} & \textbf{F1(\%)} & \textbf{Time(s)} & \textbf{ADT(s)}\\
        \midrule
        Achecker & \textbf{83.3} & \textbf{79.5} & \textbf{81.4} & 7442 & 13.5\\
        \midrule
        GPTScan & 63.1 & 66.3 & 64.7 & 8754 & 15.8\\
        \midrule
        \textbf{CKG-LLM} & 73.1 & 76.8 & 74.9 & \textbf{6248} & \textbf{11.3}\\
        \bottomrule
    \end{tabular}
\end{table}

\section{Future Plans}

In future work, we aim to enhance both the scalability and accuracy of CKG-LLM. Specifically, we will refine the prompt design and pruning strategy to support more vulnerability types, such as integer overflow or reentrancy. Additionally, we plan to incorporate retrieval-augmented generation into \cref{sec::DPO} and \cref{sec::app} by encoding contract knowledge graphs into latent embeddings, thereby reducing context overhead and guiding the LLM to generate queries compliant with Solidity grammar and contract syntax.

\begin{acks}
AI-based tools are utilized solely for language polishing during manuscript preparation.
\end{acks}

\bibliographystyle{ACM-Reference-Format}
\bibliography{acmart}

@article{shou2024llm4fuzz,
  title={Llm4fuzz: Guided fuzzing of smart contracts with large language models},
  author={Shou, Chaofan and Liu, Jing and Lu, Doudou and Sen, Koushik},
  journal={arXiv preprint arXiv:2401.11108},
  year={2024}
}

@misc{
owasp2025,
author = "OWASP",
title = "OWASP Top Ten",
howpublished = "Website",
year = {2025},
url = "https://owasp.org/Top10/2025/0x00_2025-Introduction/"
}

@article{hu2023detect,
  title={Detect defects of solidity smart contract based on the knowledge graph},
  author={Hu, Tianyuan and Li, Bixin and Pan, Zhenyu and Qian, Chen},
  journal={IEEE Transactions on Reliability},
  volume={73},
  number={1},
  pages={186--202},
  year={2023},
  publisher={IEEE}
}

@inproceedings{feist2019slither,
  title={Slither: a static analysis framework for smart contracts},
  author={Feist, Josselin and Grieco, Gustavo and Groce, Alex},
  booktitle={2019 IEEE/ACM 2nd International Workshop on Emerging Trends in Software Engineering for Blockchain (WETSEB)},
  pages={8--15},
  year={2019},
  organization={IEEE}
}

@article{chen2025forge,
  title={FORGE: An LLM-driven Framework for Large-Scale Smart Contract Vulnerability Dataset Construction},
  author={Chen, Jiachi and Shen, Yiming and Zhang, Jiashuo and Li, Zihao and Grundy, John and Shao, Zhenzhe and Wang, Yanlin and Wang, Jiashui and Chen, Ting and Zheng, Zibin},
  journal={arXiv preprint arXiv:2506.18795},
  year={2025}
}

@article{zhou2023r,
  title={R$^3$-NL2GQL: A Model Coordination and Knowledge Graph Alignment Approach for NL2GQL},
  author={Zhou, Yuhang and He, Yu and Tian, Siyu and Ni, Yuchen and Yin, Zhangyue and Liu, Xiang and Ji, Chuanjun and Liu, Sen and Qiu, Xipeng and Ye, Guangnan and others},
  journal={arXiv preprint arXiv:2311.01862},
  year={2023}
}

@inproceedings{sun2024gptscan,
  title={Gptscan: Detecting logic vulnerabilities in smart contracts by combining gpt with program analysis},
  author={Sun, Yuqiang and Wu, Daoyuan and Xue, Yue and Liu, Han and Wang, Haijun and Xu, Zhengzi and Xie, Xiaofei and Liu, Yang},
  booktitle={Proceedings of the IEEE/ACM 46th International Conference on Software Engineering},
  pages={1--13},
  year={2024}
}

@article{chen2025numscout,
  title={NumScout: Unveiling Numerical Defects in Smart Contracts using LLM-Pruning Symbolic Execution},
  author={Chen, Jiachi and Shao, Zhenzhe and Yang, Shuo and Shen, Yiming and Wang, Yanlin and Chen, Ting and Shan, Zhenyu and Zheng, Zibin},
  journal={IEEE Transactions on Software Engineering},
  year={2025},
  publisher={IEEE}
}

@misc{
soliditygrammar2025,
author = "Solidity",
title = "Language Grammar",
howpublished = "Website",
year = {2025},
url = {https://docs.soliditylang.org/en/latest/grammar.html}
}

@misc{
antlrsolidity2025,
author = "Solidity",
title = "Solidity Language Grammar",
howpublished = "Website",
year = {2025},
url = {https://github.com/solidity-parser/antlr}
}

@inproceedings{neamtiu2005understanding,
  title={Understanding source code evolution using abstract syntax tree matching},
  author={Neamtiu, Iulian and Foster, Jeffrey S and Hicks, Michael},
  booktitle={Proceedings of the 2005 international workshop on Mining software repositories},
  pages={1--5},
  year={2005}
}

@article{szabo1996smart,
  title={Smart contracts: building blocks for digital markets},
  author={Szabo, Nick},
  journal={EXTROPY: The Journal of Transhumanist Thought,(16)},
  volume={18},
  number={2},
  pages={28},
  year={1996}
}

@article{hogan2021knowledge,
  title={Knowledge graphs},
  author={Hogan, Aidan and Blomqvist, Eva and Cochez, Michael and d’Amato, Claudia and Melo, Gerard De and Gutierrez, Claudio and Kirrane, Sabrina and Gayo, Jos{\'e} Emilio Labra and Navigli, Roberto and Neumaier, Sebastian and others},
  journal={ACM Computing Surveys (Csur)},
  volume={54},
  number={4},
  pages={1--37},
  year={2021},
  publisher={ACM New York, NY, USA}
}

@inproceedings{liang2024aligning,
  title={Aligning large language models to a domain-specific graph database for nl2gql},
  author={Liang, Yuanyuan and Tan, Keren and Xie, Tingyu and Tao, Wenbiao and Wang, Siyuan and Lan, Yunshi and Qian, Weining},
  booktitle={Proceedings of the 33rd ACM International Conference on Information and Knowledge Management},
  pages={1367--1377},
  year={2024}
}

@article{chang2024survey,
  title={A survey on evaluation of large language models},
  author={Chang, Yupeng and Wang, Xu and Wang, Jindong and Wu, Yuan and Yang, Linyi and Zhu, Kaijie and Chen, Hao and Yi, Xiaoyuan and Wang, Cunxiang and Wang, Yidong and others},
  journal={ACM transactions on intelligent systems and technology},
  volume={15},
  number={3},
  pages={1--45},
  year={2024},
  publisher={ACM New York, NY}
}

@article{wei2022chain,
  title={Chain-of-thought prompting elicits reasoning in large language models},
  author={Wei, Jason and Wang, Xuezhi and Schuurmans, Dale and Bosma, Maarten and Xia, Fei and Chi, Ed and Le, Quoc V and Zhou, Denny and others},
  journal={Advances in neural information processing systems},
  volume={35},
  pages={24824--24837},
  year={2022}
}

@article{rafailov2023direct,
  title={Direct preference optimization: Your language model is secretly a reward model},
  author={Rafailov, Rafael and Sharma, Archit and Mitchell, Eric and Manning, Christopher D and Ermon, Stefano and Finn, Chelsea},
  journal={Advances in neural information processing systems},
  volume={36},
  pages={53728--53741},
  year={2023}
}

@inproceedings{ghaleb2023achecker,
  title={Achecker: Statically detecting smart contract access control vulnerabilities},
  author={Ghaleb, Asem and Rubin, Julia and Pattabiraman, Karthik},
  booktitle={2023 IEEE/ACM 45th International Conference on Software Engineering (ICSE)},
  pages={945--956},
  year={2023},
  organization={IEEE}
}

@misc{
llama,
author = "Meta",
title = "Introducing Llama 3.1: Our most capable models to date",
howpublished = "Website",
year = {2025},
url = {https://ai.meta.com/blog/meta-llama-3-1/}
}

@misc{
lcquad2,
author = "Mohnish Dubey",
title = "A large data set of natural language queries with corresponding SPARQL queries for Wikidata and Dbpedia2018",
howpublished = "Website",
year = {2025},
url = {https://github.com/AskNowQA/LC-QuAD2.0}
}

@misc{
qwen,
author = "Qwen",
title = "Alibaba Cloud's general-purpose AI models",
howpublished = "Website",
year = {2025},
url = {https://huggingface.co/Qwen/Qwen2.5-3B-Instruct}
}

@article{hu2022lora,
  title={Lora: Low-rank adaptation of large language models.},
  author={Hu, Edward J and Shen, Yelong and Wallis, Phillip and Allen-Zhu, Zeyuan and Li, Yuanzhi and Wang, Shean and Wang, Lu and Chen, Weizhu and others},
  journal={ICLR},
  volume={1},
  number={2},
  pages={3},
  year={2022}
}

@article{lewis2020retrieval,
  title={Retrieval-augmented generation for knowledge-intensive nlp tasks},
  author={Lewis, Patrick and Perez, Ethan and Piktus, Aleksandra and Petroni, Fabio and Karpukhin, Vladimir and Goyal, Naman and K{\"u}ttler, Heinrich and Lewis, Mike and Yih, Wen-tau and Rockt{\"a}schel, Tim and others},
  journal={Advances in neural information processing systems},
  volume={33},
  pages={9459--9474},
  year={2020}
}

@misc{CWE-284,
  author = {MITRE},
  title = {CWE‑284: Improper Access Control},
  howpublished = {Website},
  year = {2025},
  url = {https://cwe.mitre.org/data/definitions/284.html}
}

@misc{CWE-285,
  author = {MITRE},
  title = {CWE‑285: Improper Authorization},
  howpublished = {Website},
  year = {2025},
  url = {https://cwe.mitre.org/data/definitions/285.html}
}

@misc{CWE-863,
  author = {MITRE},
  title = {CWE‑863: Incorrect Authorization},
  howpublished = {Website},
  year = {2025},
  url = {https://cwe.mitre.org/data/definitions/863.html}
}

@misc{CWE-862,
  author = {MITRE},
  title = {CWE‑862: Missing Authorization},
  howpublished = {Website},
  year = {2025},
  url = {https://cwe.mitre.org/data/definitions/862.html}
}

@misc{CWE-269,
  author = {MITRE},
  title = {CWE‑269: Improper Privilege Management},
  howpublished = {Website},
  year = {2025},
  url = {https://cwe.mitre.org/data/definitions/269.html}
}

@misc{CWE-276,
  author = {MITRE},
  title = {CWE‑276: Incorrect Default Permissions},
  howpublished = {Website},
  year = {2025},
  url = {https://cwe.mitre.org/data/definitions/276.html}
}

@misc{CWE-732,
  author = {MITRE},
  title = {CWE‑732: Incorrect Permission Assignment for Critical Resource},
  howpublished = {Website},
  year = {2025},
  url = {https://cwe.mitre.org/data/definitions/732.html}
}

@article{li2025atomgraph,
  title={AtomGraph: Tackling Atomicity Violation in Smart Contracts using Multimodal GCNs},
  author={Li, Xiaoqi and Li, Zongwei and Li, Wenkai and Zhang, Zeng and Xie, Lei},
  journal={arXiv preprint arXiv:2512.02399},
  year={2025}
}

@article{luo2025movescanner,
  title={Movescanner: Analysis of security risks of move smart contracts},
  author={Luo, Yuhe and Li, Zhongwen and Li, Xiaoqi},
  journal={arXiv preprint arXiv:2508.17964},
  year={2025}
}

@article{li2025penetrating,
  title={Penetrating the Hostile: Detecting DeFi Protocol Exploits Through Cross-Contract Analysis},
  author={Li, Xiaoqi and Li, Wenkai and Liu, Zhiquan and Zhang, Yuqing and Mao, Yingjie},
  journal={IEEE Transactions on Information Forensics and Security},
  volume={20},
  pages={11759--11774},
  year={2025},
  publisher={IEEE}
}

@article{li2025facial,
  title={Facial Recognition Leveraging Generative Adversarial Networks},
  author={Li, Zhongwen and Li, Zongwei and Li, Xiaoqi},
  journal={arXiv preprint arXiv:2505.11884},
  year={2025}
}

@article{peng2025multicfv,
  title={Multicfv: Detecting control flow vulnerabilities in smart contracts leveraging multimodal deep learning},
  author={Peng, Hongli and Li, Xiaoqi and Li, Wenkai},
  journal={arXiv preprint arXiv:2508.01346},
  year={2025}
}

@article{gao2025implementation,
  title={Implementation and Security Analysis of Cryptocurrencies Based on Ethereum},
  author={Gao, Pengfei and Kong, Dechao and Li, Xiaoqi},
  journal={arXiv preprint arXiv:2504.21367},
  year={2025}
}

@article{zhang2025risk,
  title={Risk assessment and security analysis of large language models},
  author={Zhang, Xiaoyan and Lyu, Dongyang and Li, Xiaoqi},
  journal={arXiv preprint arXiv:2508.17329},
  year={2025}
}

@article{shen2025blockchain,
  title={When Blockchain Meets Crawlers: Real-time Market Analytics in Solana NFT Markets},
  author={Shen, Chengxin and Li, Zhongwen and Li, Xiaoqi and Li, Zongwei},
  journal={arXiv preprint arXiv:2506.02892},
  year={2025}
}

@article{wang2025ai,
  title={AI-Based Vulnerability Analysis of NFT Smart Contracts},
  author={Wang, Xin and Li, Xiaoqi},
  journal={arXiv preprint arXiv:2504.16113},
  year={2025}
}

@article{liu2025empirical,
  title={An Empirical Analysis of EOS Blockchain: Architecture, Contract, and Security},
  author={Liu, Haiyang and Mao, Yingjie and Li, Xiaoqi},
  journal={arXiv preprint arXiv:2505.15051},
  year={2025}
}

@article{peng2025mining,
  title={Mining characteristics of vulnerable smart contracts across lifecycle stages},
  author={Peng, Hongli and Li, Wenkai and Li, Xiaoqi},
  journal={IET Blockchain},
  volume={5},
  number={1},
  pages={e70016},
  year={2025},
  publisher={Wiley Online Library}
}

@article{xiang2025security,
  title={Security analysis of chatgpt: Threats and privacy risks},
  author={Xiang, Yushan and Li, Zhongwen and Li, Xiaoqi},
  journal={arXiv preprint arXiv:2508.09426},
  year={2025}
}

@article{li2025beyond,
  title={Beyond the Hype: A Large-Scale Empirical Analysis of On-Chain Transactions in NFT Scams},
  author={Li, Wenkai and Li, Zongwei and Li, Xiaoqi and Zhang, Chunyi and Zhang, Xiaoyan and Zhang, Yuqing},
  journal={arXiv preprint arXiv:2512.01577},
  year={2025}
}

@article{wu2025security,
  title={Security Vulnerabilities in Ethereum Smart Contracts: A Systematic Analysis},
  author={Wu, Jixuan and Xie, Lei and Li, Xiaoqi},
  journal={arXiv preprint arXiv:2504.05968},
  year={2025}
}

@article{zhang2025security,
  title={Security Analysis of Ponzi Schemes in Ethereum Smart Contracts},
  author={Zhang, Chunyi and Wei, Qinghong and Li, Xiaoqi},
  journal={arXiv preprint arXiv:2510.03819},
  year={2025}
}

@article{huang2025comparative,
  title={Comparative Analysis of Blockchain Systems},
  author={Huang, Jiaqi and Niu, Yuanzheng and Li, Xiaoqi and Li, Zongwei},
  journal={arXiv preprint arXiv:2505.08652},
  year={2025}
}

\end{document}